\documentclass[12pt]{article}

\newlength{\vshift}
\newlength{\hshift}
\usepackage{amssymb,amsopn,amsfonts,amsmath}
\usepackage{graphicx,latexsym}

\begin{document}

\begin{titlepage}

\vspace*{2cm}
\begin{center}
{\LARGE{\bf New Lie-Algebraic and Quadratic Deformations of Minkowski Space from Twisted Poincar\'{e} Symmetries}}

\vskip 3em

{{\bf Jerzy Lukierski\footnote{Supported by KBN grant 1P03B 01828} and Mariusz Woronowicz }

\vskip 1em
Institute for Theoretical Physics, University of Wroc{\l}aw,\\
 pl. Maxa Borna 9, 50-204 Wroc\l aw, Poland
\\
{\small{email: lukier,woronow@ift.uni.wroc.pl}}
}\\

\end{center}


\begin{abstract}
We consider two new classes of twisted D=4 quantum Poincar\'{e} symmetries described as the dual pairs of
noncocommutative Hopf algebras.
Firstly we investigate a two-parameter class of twisted Poincar\'{e} algebras which provide the examples of Lie-algebraic
noncommutativity of the translations. The corresponding associative star-products and new deformed Lie-algebraic Minkowski spaces are introduced. We discuss further the twist deformations of Poincar\'{e}
symmetries generated by the twist with its carrier in Lorentz algebra. We describe corresponding deformed Poincar\'{e} group which provides the quadratic deformations of translation sector and define the quadratically deformed Minkowski space-time algebra.
\end{abstract}

\vspace*{5.5cm}

\vfill

\end{titlepage}\vskip.2cm

\newpage
\setcounter{page}{1}
\newcommand{\Section}[1]{\setcounter{equation}{0}\section{#1}}
\renewcommand{\theequation}{\arabic{section}.\arabic{equation}}

\Section{Introduction}

In the last decade there were found arguments based on quantum gravity \cite{dfr,kma} and string theory \cite{sw,bgn},
which lead to the following noncommutative space-time coordinates (see e.g. \cite{jwess})
\begin{equation}
    \left[\hat{x}_\mu, \hat{x}_\nu\right]
    =\frac{1}{\kappa^2}\theta_{\mu\nu}\left(\kappa\hat{x}\right)\,,
    \label{rel}
\end{equation}
where $\kappa$ is a masslike parameter, and
\begin{equation}
\theta_{\mu\nu}\left(\kappa\hat{x}\right)=\theta_{\mu\nu}^{(0)}+\kappa\theta_{\mu\nu}^{(1)\  \rho}\hat{x}_\rho+\kappa^2\theta_{\mu\nu}^{(2)\ \rho\tau}\hat{x}_\rho\hat{x}_\tau+\dots\,.
\end{equation}
The introduction of noncommuting relativistic coordinate space (\ref{rel}) $\left(\theta_{\mu\nu}\neq 0\right)$ imply that
\begin{itemize}
\item[i])\,There appears in the theory the fundamental geometric mass parameter $\kappa$, which further can be identified with Planck mass.
\item[ii])\,In the theory one must introduce the collection of numerical dimensionless constant tensors $\theta_{\mu\nu}^{(k)\ \rho_1\dots\rho_k}$\,$(k=0,1,2\dots)$\,.
\end{itemize}
If we do not modify the classical relativistic (Poincar\'{e}) symmetries, these constant tensors break the Lorentz invariance.
However, one can introduce the quantum deformation of relativistic symmetries in such a way, that the particular form of the commutator (\ref{rel}) remains covariant.

The new quantum symmetry leaving invariant the simplest form of the eq. (\ref{rel}) $(\theta_{\mu\nu}=\theta_{\mu\nu}^{(0)})$ was discovered recently in \cite{jwess}-\cite{dlw5}. It has been shown that the non-vanishing commutator of space-time coordinates
\begin{equation}
\left[ \hat{x}_{\mu },\hat{x}_{\nu }\right] =\frac{i}{\kappa ^{2}}%
\,\theta _{\mu \nu }^{(0)}\,,
\label{dlww1}
\end{equation}
is covariant if we consider the Poincar\'{e} symmetries as described by a Hopf algebra and modify the Lorentz coalgebra sector through the following twist factor
\begin{equation}
\mathcal{F}_{\theta }=\exp \,\frac{i}{2\kappa^2}(\,\theta ^{\mu \nu }_{(0)}\,P_{\mu }\wedge P_{\nu
}\,).  \label{dlww2}
\end{equation}
In such a way one gets the twisted Poincar\'{e}-Hopf structure with classical Poincar\'{e} algebra relations and
modified coproducts of Lorentz generators $M_{\mu \nu }$
\begin{eqnarray}
\Delta_\theta(P_\mu)&=&\Delta_0(P_\mu), \nonumber\\
\Delta _{\theta }(M_{\mu \nu })&=&\mathcal{F}_{\theta }\circ \,\Delta _{0}(M_{\mu \nu
})\circ \mathcal{F}_{\theta }^{-1}\nonumber\\
&=&\Delta _{0}(M_{\mu \nu })-\frac{1}{\kappa ^{2}}%
\theta ^{\rho \sigma }_{(0)}[(\eta _{\rho \mu }P_{\nu }-\eta _{\rho \nu
}\,P_{\mu })\otimes P_{\sigma }\\
&&\qquad\qquad\qquad\qquad\qquad+P_{\rho}\otimes (\eta_{\sigma \mu}P_{\nu}-\eta_{\sigma \nu}P_{\mu})]\,,\nonumber\label{dlww3}
\end{eqnarray}
where $\Delta _{0}(\hat{g})=\hat{g}\otimes 1+1\otimes \hat{g}$ and $(a\otimes b)\circ
(c\otimes d)=ac\otimes bd.$

The dual $\theta_{\mu\nu}^{(0)}$-deformed Poincar\'{e} group was firstly constructed some years ago \cite{dlw6} and rediscovered recently \cite{dlw2}.
Its algebraic sector looks as follows
\begin{eqnarray}
&&\left[ \hat{a}^{\mu },\hat{a}^{\nu }\right] =-\frac{i}{\kappa^2}\,\theta ^{\rho \sigma }_{(0)}( \hat{\Lambda} _{\
\rho }^{\mu }\,\hat{\Lambda} _{\ \sigma }^{\nu }-\delta _{\ \rho }^{\mu }\,\delta
_{\ \sigma }^{\nu }) \,,  \label{dlww4}\\
&&[ \hat{\Lambda} _{\ \tau }^{\mu },\hat{\Lambda} _{\ \rho }^{\nu }] =[
\hat{a}^{\mu },\hat{\Lambda} _{\ \rho }^{\nu }] =0\,,  \label{dlws5}
\end{eqnarray}
and the coproducts remain classical
\begin{equation}
\Delta (\hat{a}^{\mu })=\hat{\Lambda} _{\ \nu }^{\mu }\otimes
\hat{a}^{\nu }+\hat{a}^{\mu }\otimes 1\,,
\qquad
\Delta (\,\hat{\Lambda} _{\ \nu }^{\mu })=\hat{\Lambda} _{\ \rho }^{\mu }\otimes \hat{\Lambda}
_{\ \nu }^{\rho }\,.  \label{dlww6}
\end{equation}
We see easily that the noncommutative space-time
(\ref{dlww1}) can be obtained by putting $\,\hat{\Lambda} _{\ \nu }^{\mu
}=0$  in the deformed translation sector (\ref{dlww4}).

The twist (\ref{dlww2}) provides an example of a "soft" or Abelian deformation%
\footnote{%
Such a deformation is sometimes called of Reshetikhin type, following twisting
procedure for Lie algebras with twist factors having the carrier in Cartan subalgebra (see
\cite{dlw8}).}, with the classical $r$-matrix
\begin{equation}
r=\frac{1}{2\kappa ^{2}}\,\theta ^{\mu \nu }_{(0)}\,P_{\mu }\wedge P_{\nu }\,,
\label{dlww7}
\end{equation}
having the Abelian carrier algebra $[P_{\mu },P_{\nu }]=0$. We recall
that the classification of all $D=4$ Poincar\'{e} algebra $r$-matrices
satisfying classical Yang-Baxter equation (CYBE) are known \cite{dlw7}.

An arbitrary $D=4$ Poincar\'{e} classical $r$-matrix can be split as follows
\begin{equation}
r=\frac{1}{2}\,\theta ^{\mu \nu \rho \sigma }_{(2)}\,M_{\mu \nu }\wedge M_{\rho \sigma
}+\frac{1}{2\kappa }\,\theta ^{\mu\nu\rho}_{(1)}\,P_{\mu }\wedge M_{\nu \rho }+%
\frac{1}{2\kappa ^{2}}\,\theta ^{\mu \nu }_{(0)}\,P_{\mu }\wedge P_{\nu }\,,
\label{dlww8}
\end{equation}
where the numerical parameters $\theta ^{\mu \nu \rho \sigma }_{(2)}$, $\theta ^{\mu\nu\rho}_{(1)}$,
$\theta ^{\mu \nu }_{(0)}$ are dimensionless and (see table 1 in \cite {dlw7} ) they are determined by the appropriate conditions.\footnote{%
For example if $\theta ^{\mu \nu \rho \sigma }_{(2)}=\theta ^{\mu\nu\rho}_{(1)}=0$ the
antisymmetric matrix $\theta ^{\mu \nu }_{(0)}$ can be arbitrary.}
\newline

In this paper we shall consider the following two classes of twist deformations:
\begin{itemize}
\item[a)] Lie-algebraic deformations.

One can choose
\begin{equation}
\theta ^{\mu\nu\rho}_{(1)}=\epsilon ^{\mu \nu \rho \tau }v_{\tau}\,,\qquad
\theta ^{\mu \nu \rho \tau }_{(2)}=\theta ^{\mu \nu }_{(0)}=0\,,  \label{dlww9}
\end{equation}
where the indices $(\nu ,\rho )$ are fixed and
$v_{\tau }$ is a numerical four-vector with two non-vanishing components.

In such a case, the classical $r$-matrix describes "soft" or Abelian
deformation with carrier algebra described by three commuting generators
$M_{\alpha\beta}\,,P_{\epsilon }, (\epsilon \neq \alpha,\,\beta;\ \alpha,\,\beta \,\,\textup{fixed})$.
The twist function is given by the formula
\begin{equation}
\mathcal{F}_{(1)}= \exp \,\frac{1}{\kappa} f_\zeta = \exp \,\frac{i}{2\kappa}( \zeta^\lambda\,P_{\lambda }\wedge M_{\alpha \beta }) \,,
\label{dlww11}
\end{equation}
where the vector $\zeta^\lambda=\theta^{\lambda\alpha\beta}_{(1)}$ has vanishing components $\zeta^\alpha,\ \zeta^\beta$.
In particular the twist (\ref{dlww11}) for the special choices $%
\{M_{\alpha\beta}= M_{03},\epsilon=1,2\}$ and $%
\{M_{\alpha\beta}=M_{12},\epsilon=0,3\}$ has been discussed some
time ago in \cite{dlw9}.

\item[b)] Quadratic deformations.

We choose
\begin{equation}
r=\frac{1}{2}\theta_{(2)}^{\alpha\beta\gamma\delta}M_{\alpha\beta}\wedge M_{\gamma\delta}\,
\end{equation}
where $\theta_{(2)}^{\alpha\beta\gamma\delta}=-\theta_{(2)}^{\beta\alpha\gamma\delta}
=-\theta_{(2)}^{\alpha\beta\delta\gamma}=-\theta_{(2)}^{\gamma\delta\alpha\beta}$ and the indices $\alpha,\,\beta\,,
\gamma\,,\delta$ are all different and fixed. The corresponding twist factor is given by the formula
\begin{equation}
    \mathcal{F}_{(2)}=\exp\,\frac{i}{2}\zeta\, M_{\alpha\beta}\wedge M_{\gamma\delta}\,,
\label{twqu}
\end{equation}
where $\zeta=\theta_{(2)}^{\alpha\beta\delta\gamma}$ is a numerical parameter.
\end{itemize}

We add here that because the carrier algebra
 for both twist functions (\ref{dlww11}) and (\ref{twqu}) is Abelian they satisfy the cocycle
   condition (see Sect. 2, Eq. (\ref{cocy})) which implies the associativity of the considered noncommutative structures.

The aim of this paper is to investigate the relativistic symmetries modified by the twists (\ref{dlww11}), (\ref{twqu})
and corresponding noncommutative Minkowski spaces.
In such a way we obtain the two-parameter family
of new deformed space-times with Lie-algebraic noncommutativity structure,
  as well as  noncommutative Minkowski spaces defined by quadratic relations.
 We observe here also that our results in the case of quadratic relations
  should be contained as isomorphic algebraic structures in general list of
  possible quantum orthogonal groups and deformed Minkowski spaces as
   their
   quantum cosets (see \cite{asch96} and \cite{asch99}, Appendix A).
   Our advantage over previous results
   which also provided link between star-products and the linear/quadratic quantum space commutators
   (see e.g. \cite{js} ) follows from more efficient starting
    point based on twisted Hopf-algebraic space-time symmetries.
     In such a way we obtain together the noncommutative space-time structure as well as the covariance of the deformed space-time algebra.

The paper is organized as follows: In Sect. 2 we describe the Hopf-algebraic form of twisted quantum Poincar\'{e}
algebra and quantum Poincar\'{e} group which are obtained by application of the twist
 (\ref{dlww11}).
In Sect. 3 we introduce corresponding associative star products and discuss the quantum Minkowski space as the representation
(Hopf algebra module) of the deformed relativistic symmetries.
In such a way the algebraic relations defining quantum Minkowski space follows as covariant under the action of twisted Poincar\'{e} algebra.
Further we shall compare the presented quantum space-time algebra with the $\kappa$-deformation providing known particular example of Lie-algebraic deformation of the Minkowski space.
In Sect. 4 we shall discuss the quadratic noncommutativity of space-time coordinates, described by $\theta^{\mu\nu\rho\sigma}_{(2)}\neq 0$.
Using the twist generated by the first term of the general formula (\ref{dlww8}) (see (\ref{twqu})) we shall define corresponding
twist-deformed Poincar\'{e} group as well as the deformed Minkowski space represented by the translation sector.
The noncommutative associative star product provides new quadratic deformation of four-dimensional space-time.
We show that the deformed Minkowski space interpreted as the translation sector of quantum Poincar\'{e} group and as the Hopf module of twisted Poincar\'{e} algebra provide the same formulae.
In Sect. 5 we shall provide an outlook. Following recent proof in \cite{wessgr} any twist satisfying cocycle condition permits to define associative noncommutative star product. One can discuss therefore along the lines of this paper also many "nonsoft" twist deformations of relativistic symmetries, described by the twist factors of Jordanian
\cite{dlw10} and generalized Jordanian \cite{dlw11,dlw12} type.

\Section{Deformed Twisted Poincar\'{e} Algebra and Poincar\'{e} Group - Lie-algebraic case}

Let us start from the classical Poincar\'{e} Hopf algebra $\mathcal{U}(%
\mathcal{P})$ described by the algebra
\begin{eqnarray}
&&\left[ M_{\mu \nu },M_{\rho \sigma }\right] =i\left( \eta _{\mu \sigma
}\,M_{\nu \rho }-\eta _{\nu \sigma }\,M_{\mu \rho }+\eta _{\nu \rho }M_{\mu
\sigma }-\eta _{\mu \rho }M_{\nu \sigma }\right) \,,  \notag \\
&&\left[ M_{\mu \nu },P_{\rho }\right] =i\left( \eta _{\nu \rho }\,P_{\mu
}-\eta _{\mu \rho }\,P_{\nu }\right) \,,  \label{dlww2.1} \\
&&\left[ P_{\mu },P_{\nu }\right] =0\,,  \notag
\end{eqnarray}
the following coalgebraic sector
\begin{eqnarray}
\Delta _{0}(P_{\mu }) =P_{\mu }\otimes 1+1\otimes P_{\mu }\,,\qquad\qquad
\Delta _{0}(M_{\mu \nu }) =M_{\mu \nu }\otimes 1+1\otimes M_{\mu \nu }\,,
 \label{dlw2.1}
\end{eqnarray}
with the counits
\begin{equation}
\epsilon(M_{\mu \nu })=\epsilon(P_\mu)=0\,,
\label{counit}
\end{equation}
and the antipodes
\begin{eqnarray}
S_{0}(P_{\mu }) =-P_{\mu }\,,  \qquad\qquad
S_{0}(M_{\mu \nu }) =-M_{\mu \nu }\,.  \label{dlw2.2}
\end{eqnarray}
We define twist $\mathcal{F}$ as an element of\, $\mathcal{U}(\mathcal{P})\otimes\mathcal{U}(\mathcal{P})$ which has an inverse,
satisfies the cocycle condition
\begin{equation}
\mathcal{F}_{12}\,\left(\Delta_0\otimes 1\right)\,\mathcal{F}=\mathcal{F}_{23}\,\left(1\otimes\Delta_0 \right)\,\mathcal{F}\,,
\label{cocy}
\end{equation}
and the normalization condition
\begin{equation}
    (\epsilon\otimes 1)\mathcal{F}=(1\otimes\epsilon)\mathcal{F}=1\,,
\end{equation}
where $\mathcal{F}_{12}=\mathcal{F}\otimes 1$ and
$\mathcal{F}_{23}=1\otimes\mathcal{F}$.
It is known, that $\mathcal{F}$ does not modify the algebraic part (\ref{dlww2.1}) and the counit, but
changes the coproducts (\ref{dlw2.1}) and the antipodes (\ref
{dlw2.2}) as follows (see e.g. \cite{dlw9})
\begin{eqnarray}
\Delta _{\mathcal{F}}(a) &=&\mathcal{F}\circ \,\Delta _{0}(a)\circ \mathcal{F%
}^{-1}\,,  \label{dlws2.1} \\
S_{\mathcal{F}}(a) &=&U\,S_{0}(a)\,U^{-1} \,,  \label{dlws2.12}
\end{eqnarray}
where due to (\ref{cocy}) the coproduct $\Delta_\mathcal{F}$ is coassociative,
$U=\sum f_{(1)}S(f_{(2)})$ (we use
Sweedler's notation $\mathcal{F}=\sum f_{(1)}\otimes f%
_{(2)}$). It should be stressed that in such a way we obtain for any choice of the twist $\mathcal{F}$ the quasitriangular Hopf algebra (see e.g. \cite{cp}).\newline

In the case of \ the twist (\ref{dlww11}), according to (\ref{dlws2.1}), one
gets
\begin{eqnarray}
\Delta(P_\mu)&=&\Delta _0(P_\mu)+(-i)^\gamma\sinh(i^{\gamma} \xi^\lambda  P_\lambda )\wedge
\left(\eta_{\alpha \mu}P_\beta -\eta_{\beta \mu}P_\alpha \right)\label{dlww2.2}\\
&+&(\cosh(i^{\gamma}\xi^\lambda  P_\lambda )-1)\perp
\left(\eta_{\alpha \alpha }\eta_{\alpha \mu}P_\alpha +\eta_{\beta \beta }\eta_{\beta \mu}P_\beta \right)\,,  \notag  \\
\Delta(M_{\mu\nu})&=&\Delta_0(M_{\mu\nu})+M_{\alpha \beta }\wedge \xi^\lambda \left(\eta_{\mu
\lambda }P_\nu-\eta_{\nu \lambda }P_\mu\right)\label{dlww2.22}\\
&+&i\left[M_{\mu\nu},M_{\alpha \beta }\right]\wedge
(-i)^{\gamma}\sinh(i^{\gamma}\xi^\lambda P_\lambda ) \notag \\
&+&\left[\left[%
M_{\mu\nu},M_{\alpha \beta }\right],M_{\alpha \beta }\right]\perp(-1)^{1+\gamma}
(\cosh(i^{\gamma}\xi^\lambda  P_\lambda  )-1)  \nonumber \\
&+&M_{\alpha \beta }(-i)^\gamma\sinh(i^{\gamma}\xi^\lambda P_\lambda )\perp
\xi^\lambda \left(\psi_\lambda P_\alpha -\chi_\lambda P_\beta \right) \notag \\
&+&\xi^\lambda \left(\psi_\lambda \eta_{\alpha \alpha }P_\beta +\chi_\lambda \eta_{\beta \beta }P_\alpha \right)\wedge
M_{\alpha \beta }(-1)^{1+\gamma}(\cosh(i^{\gamma}\xi^\lambda P_\lambda )-1)\,,
\,  \notag
\end{eqnarray}
where, in the above formulas $a\wedge b=a\otimes b-b\otimes a$, $a\perp
b=a\otimes b+b\otimes a$,
\begin{equation*}
\xi^\lambda=\frac{\zeta^\lambda}{2\kappa}\,,\qquad \psi_\lambda =\eta_{\nu \lambda }\eta_{\beta \mu}-\eta_{\mu
\lambda }\eta_{\beta \nu}\,,\qquad \chi_\lambda =\eta_{\nu \lambda }\eta_{\alpha \mu}-\eta_{\mu
\lambda }\eta_{\alpha \nu}\,,
\end{equation*}
and $\gamma=0$ when $M_{\alpha\beta}$ is a boost or $\gamma=1$ for a space rotation.
We would like to point out that for two of the momenta the coproducts remain primitive, i.e.
\begin{equation}
    \Delta(\zeta^\mu P_\mu)=\Delta_0(\zeta^\mu P_\mu)\,.
    \label{coproduct}
\end{equation}
By using (\ref{dlws2.12}) one can also check, that the antipode is classical
\begin{equation}
\qquad S_{\zeta }(P_{\mu })=-P_{\mu }\,,\qquad\qquad S_{\zeta }(M_{\mu \nu
})=-M_{\mu \nu }\,.  \label{dlww2.23}
\end{equation}

The relations (\ref{dlww2.1}), (\ref{counit}), (\ref{dlww2.2}), (\ref{dlww2.22})  and (\ref
{dlww2.23}) describe the twist-deformed Hopf Poincar\'{e} algebra $\mathcal{U}%
_{\zeta }(\mathcal{P})$ with the deformation
parameters $\xi ^{\lambda }$ proportional to the inverse of fundamental mass parameter $\kappa$.
Let us note that in the limit $\xi ^{\lambda }\rightarrow 0$
(equivalent to $\zeta\rightarrow 0\ \textup{or}\ \kappa\rightarrow \infty$) from a such deformed algebra we recover the classical limit, i.e. the Hopf structure of classical enveloping Poincar\'{e} algebra. Moreover, by straightforward calculation one
can show that $\mathcal{U}_{\zeta }(\mathcal{P})$ is real with respect the
conjugation relations $\hat{g}^*=\hat{g}$ if $(\zeta^\mu)^*=\zeta^\mu$.

The real Hopf algebra $\mathcal{U}_{\zeta }(\mathcal{P})$ is characterized by
the universal $\mathcal{R}_{(1)}$-matrix
\begin{equation}
\mathcal{R}_{(1)}=\mathcal{F}_{\zeta }^{\,T}\,\mathcal{F}_{\zeta }^{-1}=\exp (-\frac{2}{\kappa}\,f_{\zeta
})\,,\qquad \qquad (a\otimes b)^{T}=b\otimes a\,,
\label{dlww2.4}
\end{equation}
which can be used for the description of
10-generator $\zeta ^{\lambda }$-deformed Poincar\'{e} group. Using the $5\times 5$
- matrix realization of the Poincar\'{e} generators
\begin{equation}
(M_{\mu \nu })_{\ B}^{A}=\delta _{\ \mu }^{A}\eta _{\nu B }-\delta _{\
\nu }^{A}\eta _{\mu B }\,,\qquad \qquad (P_{\mu })_{\ B}^{A}=\delta
_{\ \mu }^{A}\delta _{\ B}^{4}\,,
\label{rep}
\end{equation}
we can show that in (\ref{dlww2.4}) only the term linear in $1/\kappa$ is non-vanishing
\begin{equation}
\mathcal{R}_{(1)}=1\otimes 1-\frac{i}{\kappa}\zeta^\lambda P_\lambda\wedge M_{\alpha\beta}\,.
\end{equation}

To find the matrix quantum group which is dual to the Hopf algebra $\mathcal{U}_{\zeta }(\mathcal{P})$ we introduce the following $5\times 5$ - matrices
\begin{equation}
\hat{\mathcal{T}}_{\ B}^{A}=\left(
\begin{array}{cc}
\hat{\Lambda} _{\ \nu }^{\mu } & \hat{a}^{\mu } \\
0 & 1
\end{array}
\right) \,,  \label{dlww2.7}
\end{equation}
where $\hat{\Lambda} _{\ \nu }^{\mu }$ parametrizes the quantum Lorentz rotation and $\hat{a}^{\mu }$ denotes quantum
translations. In the framework of the $\mathcal{FRT}$ procedure \cite{frt}, the algebraic sector of
such a group is described by the following relation
\begin{equation}
\mathcal{R}_{(1)}\hat{\mathcal{T}}_{1}\hat{\mathcal{T}}_{2}=\hat{\mathcal{T}}_{2}\hat{\mathcal{T}}_{1}\mathcal{R}_{(1)}\,,  \label{dlww2.8}
\end{equation}
while the composition law for the coproduct remains classical
\begin{equation}
\Delta (\hat{\mathcal{T}}_{\ B}^{A})=\hat{\mathcal{T}}_{\ C}^{A}\otimes \hat{\mathcal{T}}_{\ B}^{C}\,,  \label{dlww2.9}
\end{equation}
with $\hat{\mathcal{T}}_{1}=\hat{\mathcal{T}}\otimes 1$, $\hat{\mathcal{T}}_{2}=1\otimes \hat{\mathcal{T}}$ and quantum $\mathcal{R}_{(1)}$-matrix given in
the representation (\ref{rep}).\newline
In terms of the operator basis $(\hat{\Lambda} _{\ \nu }^{\mu },\hat{a}^{\mu })$ the
relations (\ref{dlww2.8}) can be written as follows
\begin{eqnarray}
&&\left[ \hat{a}^{\mu },\hat{a}^{\nu }\right] =\frac{i}{\kappa}\zeta ^{\nu }
( \delta _{\ \alpha }^{\mu }\hat{a}_{\beta }-\delta _{\ \beta }^{\mu }\hat{a}_{\alpha }) +
\frac{i}{\kappa}\zeta^\mu( \delta _{\ \beta }^{\nu }\hat{a}_{\alpha }-\delta _{\ \alpha }^{\nu
}\hat{a}_{\beta })  \,,  \label{dlww2.10a} \\
&&[ \hat{a}^{\mu },\hat{\Lambda} _{\ \rho }^{\nu }] =\frac{i}{\kappa}\zeta ^{\lambda }\hat{\Lambda} _{\
\lambda }^{\mu }( \eta _{\beta \rho }\hat{\Lambda} _{\ \alpha }^{\nu }-\eta _{\alpha \rho
}\hat{\Lambda} _{\ \beta }^{\nu }) +\frac{i}{\kappa}\zeta ^{\mu }( \delta _{\ \beta }^{\nu
}\hat{\Lambda} _{\alpha \rho }-\delta _{\ \alpha }^{\nu }\hat{\Lambda} _{\beta \rho }) \,,
\label{dlww2.10b} \\
&&[ \hat{\Lambda} _{\,\nu }^{\mu },\hat{\Lambda} _{\,\tau }^{\rho }] =0\,,
\label{dlww2.10c}
\end{eqnarray}
while the coproduct (\ref{dlww2.9}) takes the well known primitive form
\begin{equation}
\Delta \,(\hat{\Lambda} _{\ \nu }^{\mu })=\hat{\Lambda} _{\ \rho }^{\mu }\otimes \hat{\Lambda}
_{\ \nu }^{\rho }\,,\qquad \qquad \Delta (\hat{a}^{\mu })=\hat{\Lambda} _{\ \nu }^{\mu
}\otimes \hat{a}^{\nu }+\hat{a}^{\mu }\otimes 1\,.  \label{dlww2.11}
\end{equation}
Moreover, one can check by straightforward calculations that the above
structure together with the classical antipode
\begin{equation}
S_{\zeta }(\hat{\Lambda} _{\ \nu }^{\mu })=-\hat{\Lambda} _{\ \nu }^{\mu }\,,\qquad
\qquad S_{\zeta }(\hat{a}^{\mu })=-\hat{\Lambda} _{\ \nu }^{\mu }\,\hat{a}^{\nu }\,,
\label{dll2.12}
\end{equation}
and the counit
\begin{equation}
\epsilon(\hat{\Lambda}^\mu_{\ \nu})=\delta^\mu_{\ \nu}\,,\qquad\qquad\epsilon(\hat{a}^\mu)=0\,,
\end{equation}
define *-Hopf algebra equipped with the following *-involution
\begin{equation}
(\hat{a}^{\mu })^{\ast }=\hat{a}^{\mu }\,,\qquad \qquad (\hat{\Lambda} _{\ \nu }^{\mu })^{\ast
}=\hat{\Lambda} _{\ \nu }^{\mu }\,.  \label{dlww2.13}
\end{equation}
The relations (\ref{dlww2.10a}) - (\ref{dlww2.13}) describe the real $\zeta
^{\lambda }$-deformed Poincar\'{e} quantum group $\mathcal{G}_{\zeta }$ with noncommutative
translations $\hat{a}^{\mu }$.

\Section{New Lie-algebraic Quantum Minkowski Spaces covariant under twisted Poincar\'{e} symmetries}

In previous section we introduced the class of quantum relativistic symmetries, described by dual pairs of Hopf algebras: quantum Poincar\'{e} algebras and quantum Poincar\'{e} groups.
If we know the quantum relativistic symmetry in its Hopf algebraic form, the deformed Minkowski space can be introduced in two ways:
\begin{itemize}
\item[i])\,As the translation sector of quantum Poincar\'{e} algebra.
\item[ii])\,As the quantum representation space (a Hopf module) for quantum Poincar\'{e} algebra, with the action of the deformed symmetry generators satisfying suitably deformed Leibnitz rules \cite{dlw14b,jwess,dlw1}.
\end{itemize}
In the case of constant tensor $\theta_{\mu\nu}(\kappa\hat{x})=\theta^{(0)}_{\mu\nu}$ the translation algebra (\ref{dlww4}) is not identical with (\ref{dlww1}) - the additional $\hat{\Lambda}^\mu_\nu$-dependent terms are needed for satisfying the coproduct formulae (\ref{dlww6}) as homomorphisms. The aim of the first part of this chapter is to show that for the twist (\ref{dlww11}) these two ways of obtaining deformed Minkowski space lead to the same result.
\newline

Let us consider firstly the case of standard Poincar\'{e} symmetries acting on classical Minkowski space $\mathcal{M}$ with space-time coordinates $x_\mu$. In the case of commutative algebra $\mathcal{A}$ of functions $f(x)$ we have the representation of \ $\mathcal{U}(%
\mathcal{P})$ generated by the standard realization of \ Poincar\'{e} algebra
\begin{equation}
P_{\mu }\rhd f(x)=i\partial _{\mu }f(x)\,,\qquad \qquad M_{\mu \nu }\rhd
f(x)=i\left( x_{\mu }\partial _{\nu }-x_{\nu }\partial _{\mu }\right)
f(x)\,.  \label{a1}
\end{equation}
The action (\ref{a1}) on classical space-time coordinates looks as follows
\begin{equation}
P_{\mu }\rhd x_{\rho }=i\eta _{\mu \rho }\,,\qquad \qquad M_{\mu \nu }\rhd
x_{\rho }=i\left( x_{\mu }\eta _{\nu \rho }-x_{\nu }\eta _{\mu \rho
}\right) \,.  \label{a2}
\end{equation}
Additionally, the action of \ $\mathcal{U}(\mathcal{P})$\ on algebra $\mathcal{A}$
is consistent with the trivial coproducts (\ref{dlw2.1}), (\ref{dlw2.2}).
As a consequence, for any element $h\in \mathcal{U}(\mathcal{P})$ we have the
classical symmetric Leibniz rule
\begin{equation}
h\rhd \omega \left( f(x)\otimes g(x)\right) =\omega\circ \left( \Delta
_{0}(h)\rhd f(x)\otimes g(x)\right) \,,
\end{equation}
with the commutative multiplication $\omega$ of classical functions $\omega \circ (f(x)\otimes
g(x))=f(x)g(x)$.

The algebraic properties of the twist-deformed Minkowski space can be
derived from the consistency of new noncommutative multiplication in algebra
$\mathcal{A}$ with the twisted coproduct (\ref{dlww2.2}), (\ref{dlww2.22}%
) (see \cite{dlw14b,dlw14a}). Assuming that any element $h$ from
twisted Hopf algebra acts by deformed Leibniz rule on $\mathcal{A}$ as
follows
\begin{equation}
h\rhd \omega _{\mathcal{F}}\left( f(x)\otimes g(x)\right) =\omega _{\mathcal{%
F}}\left( \Delta _{\mathcal{F}}(h)\rhd f(x)\otimes g(x)\right)\,,
\label{dlww3.1}
\end{equation}
one can find a new noncommutative associative $\star $ - product
\begin{equation}
f(x)\star g(x):=\omega _{\mathcal{F}}\left( f(x)\otimes g(x)\right) =\omega\circ\left(
 \mathcal{F}^{-1}\rhd  f(x)\otimes g(x)\right) \,.
\label{dlww3.3}
\end{equation}
According to (\ref{a1}) and (\ref{dlww3.3}) for the twist (\ref{dlww11}) we
have
\begin{equation}
f(x)\star g(x)=\omega _{\zeta }\left( f(x)\otimes g(x)\right) =\omega\circ (
e^{\frac{i}{2\kappa}\zeta ^{\lambda }\partial _{\lambda }\wedge \left( x_{\alpha }\partial _{\beta }
-x_{\beta }\partial
_{\alpha }\right) } f(x)\otimes g(x))  \,.
\label{dlww3.4}
\end{equation}
Specifically, using (\ref{a2}) and (\ref{dlww3.4}) one can now
compute the commutator for $\zeta ^{\lambda}$-deformed Minkowski space coordinates. Because
\begin{eqnarray}
x_{\mu }\star x_{\nu } &=&x_\mu\,x_\nu
+ \frac{i}{2\kappa}\zeta _{\nu}(
\eta _{\alpha\mu }x_{\beta }-\eta _{\beta\mu }x_{\alpha }) +\frac{i}{2\kappa}\zeta_{\mu}( \eta _{
\beta\nu }x_{\alpha }-\eta _{\alpha\nu }x_{\beta })  \,, \\
x_{\nu }\star x_{\mu } &=& x_\nu\,x_\mu
- \frac{i}{2\kappa}\zeta _{\nu}(
\eta _{\alpha\mu }x_{\beta }-\eta _{\beta\mu }x_{\alpha }) -\frac{i}{2\kappa}\zeta_{\mu}( \eta _{
\beta\nu }x_{\alpha }-\eta _{\alpha\nu }x_{\beta })  \,,
\end{eqnarray}
we get
\begin{equation}
\left[ x_{\mu },x_{\nu }\right] _{\star }=x_{\mu }\star x_{\nu }-x_{\nu
}\star x_{\mu }= \frac{i}{\kappa}\zeta _{\nu}(
\eta _{\alpha\mu }x_{\beta }-\eta _{\beta\mu }x_{\alpha }) +\frac{i}{\kappa}\zeta_{\mu}( \eta _{
\beta\nu }x_{\alpha }-\eta _{\alpha\nu }x_{\beta })  \,. \label{star}
\end{equation}

The commutation relations (\ref{star}) describe the Lie-algebraic deformation
$\mathcal{M}_{\zeta}$ of the Minkowski space, which is the module (noncommutative
representation) of the Hopf algebra $\mathcal{U}_{\zeta }(\mathcal{P})$.
Comparing with (\ref{dlww2.10a}) we see that one can identify
the deformed Minkowski space with the translation sector of the quantum
group $\mathcal{G}_{\zeta }$. The relations (%
\ref{star}) can be written in more compact form as follows
\begin{equation}
\left[ x_{\mu },x_{\nu }\right] _{\star }=C_{\ \mu \nu }^{\rho }x_{\rho }\,,
\label{lie}
\end{equation}
where the constants
\begin{equation}
C_{\ \mu \nu }^{\rho }=\frac{i}{\kappa}\zeta_\mu( \eta _{\beta \nu }\delta
_{\ \alpha }^{\rho }-\eta _{\alpha \nu }\delta _{\ \beta }^{\rho })+\frac{i}{\kappa}\zeta_\nu( \eta _{\alpha \mu }\delta _{\ \beta }^{\rho }-\eta _{\beta \mu }\delta _{\ \alpha }^{\rho
}) \,,
\label{const}
\end{equation}
are the particular examples of the Lie algebra structure constants
and provide a special example of the relations (\ref{rel}).

The relation (\ref{lie}) for another choice of Lie structure constants describes also the so-called $\kappa$-deformed Minkowski space 
which in its more general form depends on the constant four-vector in the following way \cite{km}
\begin{equation}
C_{\mu\nu}^{\rho}=\frac{i}{\kappa}\zeta_\mu \delta^\rho_{\ \nu}-\frac{i}{\kappa}\zeta_\nu\delta^\rho_{\ \mu}\,.
\end{equation}
The standard choice of $\kappa$-deformation defining $\kappa$-deformed Minkowski space is obtained if $\zeta_\mu=(1,0,0,0)$.

The space-time algebra (\ref{lie}) can be written in a form recalling $\kappa$-deformed Minkowski space
relations
\begin{equation}
\left[x_\alpha,x_\lambda\right]_\star=\frac{i}{\kappa}\zeta_\lambda \eta_{\alpha\alpha}x_\beta\,,\qquad\qquad
\left[x_\beta,x_\lambda\right]_\star=-\frac{i}{\kappa}\zeta_\lambda \eta_{\beta\beta}x_\alpha\,,
\end{equation}
where $\zeta_\alpha=\zeta_\beta=0$.

Let us now provide the covariance of relations (\ref{lie}) with respect
the action of the algebra $\mathcal{U}_{\zeta }(\mathcal{P})$. According to (%
\ref{dlww3.1}) the action of $\zeta ^{\lambda }$-deformed algebra on product $%
f(x)\star g(x)$ is given by
\begin{equation}
\hat{g}\rhd \omega _{\zeta }\left( f(x)\otimes g(x)\right) =\omega _{\zeta }\left(
\Delta _{\zeta }(\hat{g})\rhd (f(x)\otimes g(x))\right) \,,  \label{dlww3.10}
\end{equation}
where $\hat{g}=(P_{\mu },M_{\mu \nu })$. In order to show the quantum covariance (\ref{lie}) we choose $\hat{g}=M_{\mu\nu}$, consider special case
$f(x)=x_\rho\,,\ g(x)=x_\sigma$ and use (\ref{dlww3.10}). In such a way, we get
\begin{eqnarray}
    M_{\mu\nu}\rhd\left(x_\rho\star x_\sigma\right)&=&\omega_\zeta\left(\Delta_\zeta\left(M_{\mu\nu}\right)\rhd\left(x_\rho\otimes x_\sigma\right)\right)   \label{act1}\\
&=&i\left(\eta_{\mu\rho}f^s_{\nu\sigma}-\eta_{\nu\rho}f^s_{\mu\sigma}+\eta_{\mu\sigma}f^s_{\nu\rho}-\eta_{\nu\sigma}f^s_{\mu\rho}\right)
+\frac{1}{2}M_{\mu\nu}\rhd C_{\ \rho\sigma}^{\lambda}x_\lambda\,,\nonumber
\end{eqnarray}
where $f^s_{\rho\sigma}=\frac{1}{2}(x_\rho\star x_\sigma+x_\sigma\star x_\rho)$ hence we have
\begin{equation}
    M_{\mu\nu}\rhd\left[x_\rho,x_\sigma\right]_\star=M_{\mu\nu}\rhd C_{\ \rho\sigma}^{\lambda}x_\lambda\,.
\end{equation}
We see therefore that the relation (\ref{lie}) is the same, with unchanged values of numerical coefficients, in any twist-deformed Lorentz frame.

\Section{New Twisted Poincar\'{e} Group and Quadratic Quantum Minkowski Spaces }

In this section we shall discuss the deformation generated by the twist (\ref{twqu}).
In analogy with (\ref{dlww2.4})
we can define the corresponding universal $\mathcal{R}_{(2)}$-matrix.
If we use the adjoint representation (\ref{rep}) one gets that the expansion
of quantum $\mathcal{R}_{(2)}$-matrix in powers of deformation parameter contains all powers of $1/\kappa$.
The closed form of the adjoint quantum $\mathcal{R}_{(2)}$-matrix looks as follows
\begin{equation}
    \mathcal{R}_{(2)}=1\otimes 1-i\sinh\zeta\,f_{\alpha_\beta\gamma\delta}+(1-\cosh\zeta)f_{\alpha\beta\gamma\delta}^2\,,
\end{equation}
where $f_{\alpha\beta\gamma\delta}\equiv M_{\alpha\beta}\wedge M_{\gamma\delta}$. In the representation (\ref{rep})
we obtain that $M_{\alpha\beta}M_{\gamma\delta}\equiv 0$, i.e.
\begin{equation}
\mathcal{R}_{(2)}=1\otimes 1-i\sinh\zeta\,M_{\alpha\beta}\wedge M_{\gamma\delta}+(1-\cosh\zeta)M_{\alpha\beta}^2\perp M_{\gamma\delta}^2\,,
\end{equation}
It can be checked that this quantum $\mathcal{R}_{(2)}$-matrix satisfy the quantum YB equation
\begin{equation}
    \mathcal{R}_{12}\mathcal{R}_{13}\mathcal{R}_{23}=\mathcal{R}_{23}\mathcal{R}_{13}\mathcal{R}_{12}\,,
\end{equation}
where if $\mathcal{R}=\mathcal{R}_\alpha\otimes\mathcal{R}_\beta$ then
$\mathcal{R}_{12}=\mathcal{R}_\alpha\otimes\mathcal{R}_\beta\otimes 1$ etc.

Again we apply the $\mathcal{FRT}$ procedure \cite{frt}. The relation
\begin{equation}
    \mathcal{R}_{(2)}\hat{\mathcal{T}}_1\hat{\mathcal{T}}_2=\hat{\mathcal{T}}_2\hat{\mathcal{T}}_1\mathcal{R}_{(2)}\,,
\end{equation}
leads to the following algebraic sector of the twist-deformed matrix Poincar\'{e} group (see (\ref{dlww2.7}))
\begin{eqnarray}
[\hat{\mathcal{T}}^A_{\ P}, \hat{\mathcal{T}}^B_{\ Q}]&=&
(1-\cosh\zeta)
\sum_{\genfrac{}{}{0pt}{}{k=\alpha,\beta}{l=\gamma,\delta}}^{}
(\delta^{A}_ {\ \{k}\delta^{B}_{\ l\}}\hat{\mathcal{T}}^{\{k}_{\ P} \hat{\mathcal{T}}^{l\}}_{\ Q}
-\delta^{\{k}_ {\ Q}\delta^{l\}}_{\ P}\hat{\mathcal{T}}^B_{\ \{k} \hat{\mathcal{T}}^A_{\ l\}})\label{qrtt}\\
&+&i\sinh\zeta[(\eta_{\beta Q}\hat{\mathcal{T}}^B_{\ \alpha}-\eta_{\alpha Q}\hat{\mathcal{T}}^B_{\ \beta})
(\eta_{\delta P}\hat{\mathcal{T}}^A_{\ \gamma}-\eta_{\gamma P}\hat{\mathcal{T}}^A_{\ \delta}) \nonumber\\
&&\qquad\qquad\qquad+(\eta_{\delta Q}\hat{\mathcal{T}}^B_{\ \gamma}-\eta_{\gamma Q}\hat{\mathcal{T}}^B_{\ \delta})
(\eta_{\alpha P}\hat{\mathcal{T}}^A_{\ \beta}-\eta_{\beta P}\hat{\mathcal{T}}^A_{\ \alpha})\nonumber\\
&&\qquad\qquad\qquad+(\eta^{\beta A}\hat{\mathcal{T}}^\alpha_{\ P}-\eta^{\alpha A}\hat{\mathcal{T}}^\beta_{\ P})
(\eta^{\gamma B}\hat{\mathcal{T}}^\delta_{\ Q}-\eta^{\delta B}\hat{\mathcal{T}}^\gamma_{\ Q})\nonumber\\
&&\qquad\qquad\qquad+(\eta^{\gamma A}\hat{\mathcal{T}}^\delta_{\ P}-\eta^{\delta A}\hat{\mathcal{T}}^\gamma_{\ P})
(\eta^{\alpha B}\hat{\mathcal{T}}^\beta_{\ Q}-\eta^{\beta B}\hat{\mathcal{T}}^\alpha_{\ Q})]\,,\nonumber
\end{eqnarray}
where $\delta^{A}_ {\ \{k}\delta^{B}_{\ l\}}\hat{\mathcal{T}}^{\{k}_{\ P} \hat{\mathcal{T}}^{l\}}_{\ Q}
=\delta^{A}_ {\ k}\delta^{B}_{\ l}\hat{\mathcal{T}}^{k}_{\ P} \hat{\mathcal{T}}^{l}_{\ Q}+
\delta^{A}_ {\ l}\delta^{B}_{\ k}\hat{\mathcal{T}}^{l}_{\ P} \hat{\mathcal{T}}^{k}_{\ Q}$.
Putting $\hat{\mathcal{T}}^\mu_{\ 4}\equiv \hat{a}^\mu$ i $\hat{\mathcal{T}}^\nu_{\ \rho}\equiv\hat{\Lambda}^\nu_{\ \rho}$ we get the relations (\ref{qrtt}) in more explicit form
\begin{eqnarray}
[\hat{a}^\mu, \hat{a}^\nu]&=&(1-\cosh\zeta)
\sum_{\genfrac{}{}{0pt}{}{k=\alpha,\beta}{l=\gamma,\delta}}^{}
\delta^{\mu}_ {\ \{k}\delta^{\nu}_{\ l\}}\hat{a}^{\{k} \hat{a}^{l\}}
\label{quadratic}\\
    &+&i\sinh\zeta[(\delta^{\mu}_{\ \alpha}\hat{a}_\beta-\delta^{\mu}_{\ \beta}\hat{a}_\alpha)
(\delta^{\nu}_{\ \gamma}\hat{a}_\delta-\delta^{\nu}_{\ \delta}\hat{a}_\gamma)\nonumber\\
&&\qquad\qquad\qquad-(\delta^{\mu}_{\ \gamma}\hat{a}_\delta-\delta^{\mu}_{\ \delta}\hat{a}_\gamma)
(\delta^{\nu}_{\ \alpha}\hat{a}_\beta-\delta^{\nu}_{\ \beta}\hat{a}_\alpha)]\,,\nonumber
\end{eqnarray}
or equivalently
\begin{eqnarray}
[\hat{a}^\mu, \hat{a}^\nu]_q&=&i\sinh\zeta[(\delta^{\mu}_{\ \alpha}\hat{a}_\beta-\delta^{\mu}_{\ \beta}\hat{a}_\alpha)
(\delta^{\nu}_{\ \gamma}\hat{a}_\delta-\delta^{\nu}_{\ \delta}\hat{a}_\gamma)\label{cqc}\\
&&\qquad\qquad\qquad-(\delta^{\mu}_{\ \gamma}\hat{a}_\delta-\delta^{\mu}_{\ \delta}\hat{a}_\gamma)
(\delta^{\nu}_{\ \alpha}\hat{a}_\beta-\delta^{\nu}_{\ \beta}\hat{a}_\alpha)]\,,\nonumber
\end{eqnarray}
where $[\hat{a}^\mu, \hat{a}^\nu]_q\equiv\hat{a}^\mu \hat{a}^\nu -\hat{a}^\nu\hat{a}^\mu
+ (\cosh\zeta-1)
\sum_{\genfrac{}{}{0pt}{}{k=\alpha,\beta}{l=\gamma,\delta}}^{}
\delta^{\mu}_ {\ \{k}\delta^{\nu}_{\ l\}}
\hat{a}^{\{k} \hat{a}^{l\}}$.
Additionally
\begin{eqnarray}
[\hat{a}^\mu, \hat{\Lambda}^\nu_{\ \rho}]&=&(1-\cosh\zeta)
\sum_{\genfrac{}{}{0pt}{}{k=\alpha,\beta}{l=\gamma,\delta}}^{}
\delta^{\mu}_ {\ \{k}\delta^{\nu}_{\ l\}}\hat{a}^{\{k} \hat{\Lambda}_{\ \rho}^{l\}}\\
    &+&i\sinh\zeta[(\delta^{\mu}_{\ \alpha}\hat{a}_\beta-\delta^{\mu}_{\ \beta}\hat{a}_\alpha)
(\delta^{\nu}_{\ \gamma}\hat{\Lambda}_{\delta \rho}-\delta^{\nu}_{\ \delta}\hat{\Lambda}_{\gamma \rho})\nonumber\\
&&\qquad\qquad\qquad-(\delta^{\mu}_{\ \gamma}\hat{a}_\delta-\delta^{\mu}_{\ \delta}\hat{a}_\gamma)
(\delta^{\nu}_{\ \alpha}\hat{\Lambda}_{\beta \rho}-\delta^{\nu}_{\ \beta}\hat{\Lambda}_{\alpha \rho})]\,,\nonumber
\end{eqnarray}
and
\begin{eqnarray}
[\hat{\Lambda}^\mu_{\ \tau},\hat{\Lambda}^\nu_{\ \rho}]&=&[\hat{\mathcal{T}}^\mu_{\ \tau}, \hat{\mathcal{T}}^\nu_{\ \rho}]_{\mid \hat{\mathcal{T}}\rightarrow \hat{\Lambda}}\,,
\label{quadratic1}
\end{eqnarray}
with explicit form of rhs of (\ref{quadratic1}) given by rhs of (\ref{qrtt}) with all five-dimensional indices $A,\ B,\dots$ replaced by four-dimensional ones $\mu,\ \nu,\dots$.
We see that the generators $M_{\alpha\beta},\ M_{\gamma\delta}$, which describe the twist (\ref{twqu}), distinguish two sets of indices
 $\mathcal{A}=\{\alpha,\ \beta\}$ and $\mathcal{B}=\{\gamma,\ \delta\}$.
 One obtains
$[\hat{\mathcal{T}}^\mu_{\ \tau},\hat{\mathcal{T}}^\nu_{\ \rho}]\neq 0$ only if at least one pair of upper or lower indices
comes from different sets $\mathcal{A},\,\mathcal{B}$.

For our quantum group the coproduct, the counit and the antipode remain classical.
The relations (\ref{quadratic}) - (\ref{quadratic1}) describe the real $\zeta$-deformed Poincar\'{e} group
with ${}^*$-involution, $(\hat{a}^{\mu })^{\ast }=\hat{a}^{\mu },\ (\hat{\Lambda} _{\ \nu }^{\mu })^{\ast
}=\hat{\Lambda} _{\ \nu }^{\mu }$ if
 $ \    \zeta^*=-\zeta\,.
$

Subsequently we shall consider the deformed Minkowski space as Hopf algebra module of the twist-deformed relativistic symmetries. Using the general formula (\ref{dlww3.3}) we obtain the following $\star$ - product (let us recall that $\omega\circ (a\otimes b)=ab$)
\begin{eqnarray}
x_\mu\star x_\nu&=&\omega\circ (e^{\frac{i}{2}\zeta(x_\beta\partial_\alpha-x_\alpha\partial_\beta)\wedge
(x_\delta\partial_\gamma-x_\gamma\partial_\delta)}x_\mu\otimes x_\nu)\label{starq} \\
&=&\omega\circ x_\mu \otimes x_\nu
+i\sinh\frac{\zeta}{2} \omega\circ[
(x_\beta\eta_{\alpha\mu}-x_\alpha\eta_{\beta\mu})\otimes
(x_\delta\eta_{\gamma\nu}-x_\gamma\eta_{\delta_\nu})\nonumber\\
&&\qquad\qquad\qquad\qquad\qquad\qquad-(x_\delta\eta_{\gamma\mu}-x_\gamma\eta_{\delta\mu})\otimes(x_\beta\eta_{\alpha\nu}-x_\alpha\eta_{\beta\nu})]\nonumber\\
&+&(\cosh\frac{\zeta}{2}-1)\omega\circ[\sum_{\genfrac{}{}{0pt}{}{k=\alpha,\beta}{l=\gamma,\delta}}^{}
(x_{k}\delta^{k}_{\ \mu}\otimes x_{l}\delta^l_{\ \nu}+x_{l}\delta^{l}_{\ \mu}\otimes x_{k}\delta^{k}_{\ \nu})]\nonumber\,,\\
x_\nu\star x_\mu&=&x_\mu\star x_\nu \ _{\mid \mu\leftrightarrow \nu}\nonumber\,,
\end{eqnarray}
hence
\begin{eqnarray}
[x_\mu,x_\nu]_\star&=&\omega\circ[x_\mu,x_\nu]_\otimes\label{qsp}\\
&+&i\sinh\frac{\zeta}{2} \omega\circ[
(\eta_{\alpha[\mu}\eta_{\gamma\nu]}\{x_\beta,x_\delta\}_\otimes
-\eta_{\alpha[\mu}\eta_{\delta\nu]}\{x_\beta,x_\gamma\}_\otimes\nonumber\\
&&\qquad\qquad\qquad-
\eta_{\beta[\mu}\eta_{\gamma\nu]}\{x_\alpha,x_\delta\}_\otimes
+\eta_{\beta[\mu}\eta_{\delta\nu]}\{x_\alpha,x_\gamma\}_\otimes]\nonumber\\
&+&(\cosh\frac{\zeta}{2}-1)\omega\circ(\sum_{\genfrac{}{}{0pt}{}{k=\alpha,\beta}{l=\gamma,\delta}}^{}
\delta^k_{\ [\mu}\delta^l_{\ \nu]}[x_k,x_l]_\otimes)\nonumber\\
&=&i\sinh\frac{\zeta}{2} [
\eta_{\alpha[\mu}\eta_{\gamma\nu]}\{x_\beta,x_\delta\}
-\eta_{\alpha[\mu}\eta_{\delta\nu]}\{x_\beta,x_\gamma\}\nonumber\\
&&\qquad\qquad\qquad-
\eta_{\beta[\mu}\eta_{\gamma\nu]}\{x_\alpha,x_\delta\}
+\eta_{\beta[\mu}\eta_{\delta\nu]}\{x_\alpha,x_\gamma\}]\,,\nonumber
\end{eqnarray}
where $\eta_{\alpha[\mu}\eta_{\gamma\nu]}=\eta_{\alpha\mu}\eta_{\gamma\nu}-\eta_{\alpha\nu}\eta_{\gamma\mu}$.
In the formula (\ref{qsp}) on rhs there is used the standard Abelian multiplication of the functions of classical Minkowski space. If we use the formula inverse to the one given in (\ref{starq}) expressing
standard product in terms of star products (see e.g. \cite{asd})
\begin{equation}
    x_\mu x_\nu=\omega\circ(\mathcal{F}^{-1}\mathcal{F}x_\mu\otimes x_\nu)=\omega_\mathcal{F}\circ (e^{-\frac{i}{2}\zeta(x_\beta\partial_\alpha-x_\alpha\partial_\beta)\wedge
(x_\delta\partial_\gamma-x_\gamma\partial_\delta)}x_\mu\otimes x_\nu)\,,
\end{equation}
one obtains the following form of the relation (\ref{qsp})
\begin{eqnarray}
[x_\mu,x_\nu]_\star&=&i\sinh\frac{\zeta}{2}\cosh\frac{\zeta}{2}
(\eta_{\alpha[\mu}\eta_{\gamma\nu]}\{x_\beta,x_\delta\}_\star
-\eta_{\alpha[\mu}\eta_{\delta\nu]}\{x_\beta,x_\gamma\}_\star\\
&&\qquad\qquad\qquad\qquad-
\eta_{\beta[\mu}\eta_{\gamma\nu]}\{x_\alpha,x_\delta\}_\star
+\eta_{\beta[\mu}\eta_{\delta\nu]}\{x_\alpha,x_\gamma\}_\star)\nonumber\\
&-&\sinh^2\frac{\zeta}{2}(\sum_{\genfrac{}{}{0pt}{}{k=\alpha,\beta}{l=\gamma,\delta}}^{}
\delta^k_{\ [\mu}\delta^l_{\ \nu]}[x_k,x_l]_\star)\,.\nonumber
\end{eqnarray}
We may rewrite translation sector (\ref{quadratic}) in the form
\begin{eqnarray}
    [\hat{a}^\mu,\hat{a}^\nu]&=&\frac{i}{2}\sinh\zeta(
\delta^{[\mu}_{\ \alpha}\delta^{\nu]}_{\ \gamma}\{\hat{a}_\beta,\hat{a}_\delta\}
-\delta^{[\mu}_{\ \alpha}\delta^{\nu]}_{\ \delta}\{\hat{a}_{\beta},\hat{a}_\gamma\}\\
&&\qquad\qquad-\delta^{[\mu}_{\ \beta}\delta^{\nu]}_{\ \gamma}\{\hat{a}_\alpha,\hat{a}_\delta\}
+\delta^{[\mu}_{\ \beta}\delta^{\nu]}_{\ \delta}\{\hat{a}_\alpha,\hat{a}_\gamma\})\nonumber\\
&+&\frac{1}{2}(1-\cosh\zeta)\sum_{\genfrac{}{}{0pt}{}{k=\alpha,\beta}{l=\gamma,\delta}}^{}
\delta_{\ k}^{[\mu}\delta_{\ l}^{\nu]}[\hat{a}^k,\hat{a}^l]\,.\nonumber
\end{eqnarray}
Putting $k\in\mathcal{A}$ and $l\in\mathcal{B}$ one gets the following non-vanishing commutators
\begin{eqnarray}
&&[x_k,x_l]_\star=i\tanh\frac{\zeta}{2}
(\eta_{\alpha k}\eta_{\gamma l}\{x_\beta,x_\delta\}_\star
-\eta_{\alpha k}\eta_{\delta l}\{x_\beta,x_\gamma\}_\star\label{star1q}\\
&&\qquad\qquad\qquad\qquad-
\eta_{\beta k}\eta_{\gamma l}\{x_\alpha,x_\delta\}_\star
+\eta_{\beta k}\eta_{\delta l}\{x_\alpha,x_\gamma\}_\star)\,,\nonumber\\
&&[\hat{a}^k,\hat{a}^l]=i\tanh\frac{\zeta}{2}(
\delta^{k}_{\ \alpha}\delta^{l}_{\ \gamma}\{\hat{a}_\beta,\hat{a}_\delta\}
-\delta^{k}_{\ \alpha}\delta^{l}_{\ \delta}\{\hat{a}_{\beta},\hat{a}_\gamma\}\label{star2q}\\
&&\qquad\qquad\qquad\qquad-\delta^{k}_{\ \beta}\delta^{l}_{\ \gamma}\{\hat{a}_\alpha,\hat{a}_\delta\}
+\delta^{k}_{\ \beta}\delta^{l}_{\ \delta}\{\hat{a}_\alpha,\hat{a}_\gamma\})\,,\nonumber
\end{eqnarray}
which are the same for Minkowski space generators describing the Hopf module of twisted Poincar\'{e} algebra and for the translation sector of the corresponding quantum Poincar\'{e} group.
One can add that both multiplications in the formulae (\ref{star1q}), (\ref{star2q}) are associative which is guaranteed by the cocycle condition (\ref{cocy}) \cite{wessgr}.

\Section{Outlook}

Introduction of quantum symmetries as described by Hopf algebras permits to preserve the noncommutative structures depending on constant tensors and central elements. Particularly efficient is the description of quantum groups generated by the classical r-matrices satisfying classical Yang-Baxter algebra. It appears that in such a case the quantization of the symmetry is due to the twist factor, modifying the coalgebra structure. If $\mathcal{A}$  is the quantum symmetry module, the twist $\mathcal{F}$ with the tensor structure $\mathcal{F}^{-1}=\overline{f}_{(1)}\otimes\overline{f}_{(2)}$ determines the noncommutative multiplication in $\mathcal{A}$ $(a,\  b\in\mathcal{A})$
\begin{equation}
    \omega_\mathcal{F}(a\cdot b)=(\overline{f}_{(1)}a)(\overline{f}_{(2)}b)\,.
\end{equation}
The novelty of this paper is to consider in physical basis
 some explicit examples of twists of Poincar\'{e} algebra which provide linear and quadratic space-time commutators.

In this paper we consider only the Abelian twists, with commutative carrier algebras. One can also introduce graded-Abelian twists, applied to super-Poincar\'{e} algebra, and obtain the covariance of deformed super-space relations \cite{ksasaki}.
The analogous discussion can be applied as well to the
non-Abelian twists, in particular to the simple Jordanian twist for $SL(2,\mathbb{C})$ \cite{dlw10} and its real forms.
Such twists were classified by S. Zakrzewski \cite{dlw14} and now are under consideration.
The next step is to study the twisted quantum symmetries and star product multiplication for quantum representation spaces in the case of extended twists
(see e.g. \cite{dlw11,dlw12}). In particular the noncommutative representation spaces (e.g. quantum six-dimensional conformal
space-time with signature (++++$--$)) can be studied in the case of twisted $D=4$ conformal symmetries \cite{dlw13}.

The twisted quantum geometries permit also to consider in explicit way as the representation spaces the quantum fields on noncommutative space-time and consider the noncommutative field theory with quantum group invariance.
Following recent result of Wess and all \cite{asd} one can also apply the twist deformation to infinite-dimensional coalgebra of general coordinate transformations and deduce the definitions of the basic geometric objects in gravity theory on noncommutative space-time.

\paragraph{Acknowledgements}
We would like to thank Marcin Daszkiewicz for many valuable discussions and Paolo Aschieri for pointing
  out ref. \cite{asch96,asch99}.
We are also grateful to Piotr Kosi\'{n}ski and Pawe\l\ Ma\'{s}lanka for an important algebraic remark.


\begin{thebibliography}{99}

\bibitem{dfr}  S. Doplicher, K. Fredenhagen and J.E. Roberts, Phys. Lett. \textbf{B331}, 39 (1994);
Commun. Math. Phys. 172, 187 (1995); hep-th/0303037.

\bibitem{kma}  A. Kempf and G. Mangano, Phys. Rev. \textbf{D55}, 7909 (1997); hep-th/9612084.

\bibitem{sw}  N. Seiberg and E. Witten, JHEP \textbf{09}, 032 (1999); hep-th/9908142.

\bibitem{bgn} J. de Boer, P. A. Grassi and P. van Nieuwenhuizen, Phys. Lett. \textbf{B574}, 98 (2003); hep-th/0302078.

\bibitem{jwess} J. Wess, \textit{Deformed coordinate spaces: Derivatives}; hep-th/0408080.

\bibitem{dlw1}  M. Chaichian, P.P Kulish, K. Nishijima and A. Tureanu, Phys.
Lett. \textbf{B604}, 98--102 (2004); hep-th/0408069.

\bibitem{dlw2}  P. Kosi\'{n}ski and P. Ma\'{s}lanka, \textit{Lorentz - invariant interpretation of noncommutative space-time: Global version}; hep-th/0408100.

\bibitem{dlw3}  F. Koch and E. Tsouchnika, Nucl. Phys. \textbf{B717}, 387 (2005); hep-th/0409012.

\bibitem{dlw4}  M. Chaichian, P. Presnajder and A. Tureanu, Phys. Rev. Lett. \textbf{94}, 151602 (2005); hep-th/0409096.

\bibitem{dlw5}  C. Gonera, P. Kosinski, P. Maslanka and S. Giller, \textit{Space-time symmetry of noncommutative field theory}; hep-th/0504132.

\bibitem{dlw6}  R. Oeckl, Nucl. Phys. \textbf{B581}, 559 (2000); hep-th/0003018.

\bibitem{dlw7}  S. Zakrzewski, \textit{Poisson Poincar\'{e} groups}; hep-th/9412099.

\bibitem{dlw8}  N.Yu. Reshetikhin, Lett. Math. Phys. \textbf{20}, 331 (1990).

\bibitem{dlw9}  J. Lukierski, A. Nowicki, H. Ruegg and V.N. Tolstoy, J.
Phys. \textbf{A27}, 2389 (1994); hep-th/9312068.



\bibitem{asch96} P. Aschieri and L. Leonardo, Int. J. Mod. Phys. {\bf A11}, 4513 (1996); q-alg/9601006.

\bibitem{asch99} P. Aschieri, L. Castellani and A.M. Scarfone, Eur. Phys. J. {\bf C7},
 159 (1999).

\bibitem{js} C. Jambor and A. Sykora, \textit{Realization of algebras with the help of $\star$-products}; hep-th/0405268.

\bibitem{wessgr} P. Aschieri, M. Dimitrijevic, F. Meyer and J. Wess \textit{Noncommutative geometry and gravity}; hep-th/0510059.

\bibitem{dlw10}  V. Ogievetsky, Suppl. Rendiconti Gr. Math. Palermo, Serie II \textbf{37}, 185 (1993).

\bibitem{dlw11}  P.P. Kulish, V.D. Lyakhovsky and A. Mudrov, J. Math. Phys. \textbf{40}, 4569 (1993).

\bibitem{dlw12}  P.P. Kulish, V.D. Lyakhovsky and M.A. del Olmo, J. Phys. \textbf{A32}, 8671 (1999).

\bibitem{cp}  V. Chari and A. Pressley, \textit{A Guide to Quantum Groups}, Cambridge: University Press (1995).

\bibitem{frt} L. D. Faddeev, N. Yu. Reshetikhin and L. A. Takhtadzhyan, Leningrad Math. J. \textbf{1} (1990).

\bibitem{dlw14b} C. Blohmann, J. Math. Phys. \textbf{44}, 4736 (2003); q-alg/0209180.

\bibitem{dlw14a} C. Blohmann, \textit{Realization of q deformed space-time as star product by a Drinfeld twist}; q-alg/0402199.

\bibitem{km} P. Kosi\'{n}ski and P. Ma\'{s}lanka, in "From Quantum Field Theory to Quantum Groups", ed. B. Jancewicz and J. Sobczyk, World Scientific, p. 41, q-alg/9606022.

\bibitem{asd}  P. Aschieri, Ch. Blohmann, M. Dimitrijevic, F. Meyer, P. Schupp  and J. Wess, \textit{A gravity theory on noncommutative spaces}; hep-th/0504183.

\bibitem{ksasaki} Y. Kobayashi, S. Sasaki, \textit{Lorentz invariant and supersymmetric interpretation of noncommutative quantum field theory}; hep-th/0410164.

\bibitem{dlw14}  S. Zakrzewski, Lett. Math. Phys. \textbf{32}, 11 (1994).

\bibitem{dlw13}  J. Lukierski, V.D. Lyakhovsky and M. Mozrzymas, Phys. Lett. \textbf{B538}, 375 (2002); hep-th/0203182.

\end{thebibliography}
\end{document}